\documentclass[aip,jcp,reprint,noshowkeys,superscriptaddress]{revtex4-1}
\usepackage{graphicx,dcolumn,bm,xcolor,microtype,multirow,amscd,amsmath,amssymb,amsfonts,physics,longtable,wrapfig,txfonts,siunitx}
\usepackage[version=4]{mhchem}

\usepackage[utf8]{inputenc}
\usepackage[T1]{fontenc}
\usepackage{txfonts}

\usepackage[
	colorlinks=true,
    citecolor=blue,
    breaklinks=true
	]{hyperref}
\newcommand{\ie}{\textit{i.e.}}

\newcommand{\alert}[1]{\textcolor{black}{#1}}
\usepackage[normalem]{ulem}

\newcommand{\T}[1]{#1^{\intercal}}

\newcommand{\br}{\boldsymbol{r}}
\newcommand{\bx}{\boldsymbol{x}}

\newcommand{\GW}{GW}

\newcommand{\dRPA}{\text{dRPA}}

\newcommand{\BSE}{\text{BSE}}

\newcommand{\hH}{\Hat{H}}

\newcommand{\bHN}{\Bar{H}_{\text{N}}}
\newcommand{\tHN}{\Tilde{H}_{\text{N}}}
\newcommand{\hT}{\Hat{T}}

\newcommand{\cF}{\mathcal{F}}
\newcommand{\cW}{\mathcal{W}}

\newcommand{\Ec}{E_\text{c}}

\newcommand{\ECC}{E_\text{CC}}

\newcommand{\e}[2]{\epsilon_{#1}^{#2}}

\newcommand{\Om}[2]{\Omega_{#1}^{#2}}

\newcommand{\SO}[2]{\psi_{#1}^{#2}}
\newcommand{\ERI}[2]{\braket{#1}{#2}}
\newcommand{\sERI}[2]{(#1|#2)}
\newcommand{\dbERI}[2]{\mel{#1}{}{#2}}
\newcommand{\wERI}[2]{\widetilde{W}_{#1 #2}}

\newcommand{\Sig}[2]{\Sigma_{#1}^{#2}}

\newcommand{\bO}{\boldsymbol{0}}
\newcommand{\bI}{\boldsymbol{1}}

\newcommand{\bOm}[2]{\boldsymbol{\Omega}_{#1}^{#2}}

\newcommand{\bA}[2]{\boldsymbol{A}_{#1}^{#2}}
\newcommand{\bB}[2]{\boldsymbol{B}_{#1}^{#2}}
\newcommand{\bC}[2]{\boldsymbol{C}_{#1}^{#2}}

\newcommand{\bR}[2]{\boldsymbol{R}_{#1}^{#2}}
\newcommand{\bT}[2]{\boldsymbol{T}_{#1}^{#2}}

\newcommand{\bV}[2]{\boldsymbol{V}_{#1}^{#2}}
\newcommand{\bX}[2]{\boldsymbol{X}_{#1}^{#2}}
\newcommand{\bY}[2]{\boldsymbol{Y}_{#1}^{#2}}

\newcommand{\be}[2]{\boldsymbol{\epsilon}_{#1}^{#2}}
\newcommand{\bSig}[2]{\boldsymbol{\Sigma}_{#1}^{#2}}

\newcommand{\ii}{\mathrm{i}}

\newcommand{\LCPQ}{Laboratoire de Chimie et Physique Quantiques (UMR 5626), Universit\'e de Toulouse, CNRS, UPS, France}

\begin{document}	

\title{Connections between many-body perturbation and coupled-cluster theories}

\author{Ra\'ul \surname{Quintero-Monsebaiz}}
	\affiliation{\LCPQ}
\author{Enzo \surname{Monino}}
	\affiliation{\LCPQ}
\author{Antoine \surname{Marie}}
	\affiliation{\LCPQ}
\author{Pierre-Fran\c{c}ois \surname{Loos}}
	\email{loos@irsamc.ups-tlse.fr}
	\affiliation{\LCPQ}

\begin{abstract}
Here, we build on the works of Scuseria \textit{et al.} [\href{http://dx.doi.org/10.1063/1.3043729}{J.~Chem.~Phys.~\textbf{129}, 231101 (2008)}] and Berkelbach [\href{https://doi.org/10.1063/1.5032314}{J.~Chem.~Phys.~\textbf{149}, 041103 (2018)}] to show connections between the Bethe-Salpeter equation (BSE) formalism combined with the $GW$ approximation from many-body perturbation theory and coupled-cluster (CC) theory at the ground- and excited-state levels.
In particular, we show how to recast the $GW$ and Bethe-Salpeter equations as non-linear CC-like equations.
Similitudes between BSE@$GW$ and the similarity-transformed equation-of-motion CC method introduced by Nooijen are also put forward.
The present work allows to easily transfer key developments and general knowledge gathered in CC theory to many-body perturbation theory.
In particular, it may provide a path for the computation of ground- and excited-state properties (such as nuclear gradients) within the $GW$ and BSE frameworks.
\end{abstract}

\maketitle

\section{RPA Physics and Beyond}
The random-phase approximation (RPA), introduced by Bohm and Pines \cite{Bohm_1951,Pines_1952,Bohm_1953} in the context of the uniform electron gas, \cite{Loos_2016} is a quasibosonic approximation where one treats fermion products as bosons.
In the particle-hole (ph) channel, which is quite popular in the electronic structure community, \cite{Ren_2012,Chen_2017} particle-hole fermionic excitations and deexcitations are assumed to be bosons. 
Because ph-RPA takes into account dynamical screening by summing up to infinity the (time-independent) ring diagrams, it is adequate in the high-density (or weakly correlated) regime and captures effectively long-range correlation effects (such as dispersion). \cite{Gell-Mann_1957,Nozieres_1958} 
Another important feature of ph-RPA compared to finite-order perturbation theory is that it does not exhibit divergences for small-gap or metallic systems. \cite{Gell-Mann_1957}

Roughly speaking, the Bethe-Salpeter equation (BSE) formalism \cite{Salpeter_1951,Strinati_1988,Blase_2018,Blase_2020} of many-body perturbation theory \cite{Martin_2016} can be seen as a cheap and efficient way of introducing correlation in order to go \textit{beyond} RPA physics.
In the ph channel, BSE is commonly performed on top of a $GW$ calculation \cite{Hedin_1965,Aryasetiawan_1998,Onida_2002,Reining_2017,Golze_2019,Bruneval_2021} from which one extracts the quasiparticle energies as well as the dynamically-screened Coulomb potential $W$. 
Practically, $GW$ produces accurate \textit{``charged''} excitations providing a faithful description of the fundamental gap via the computation of the RPA polarizability obtained by a resummation of all time-dependent ring diagrams. 
The remaining excitonic effect (\ie, the stabilization provided by the attraction of the excited electron and its hole left behind) is caught via BSE, hence producing overall accurate \textit{``neutral''} excitations.
BSE@$GW$ has been shown to be highly successful to compute low-lying excited states of various natures (charge transfer, Rydberg, valence, etc) in molecular systems with a very attractive accuracy/cost ratio.\cite{Rohlfing_1999a,Horst_1999,Puschnig_2002,Tiago_2003,Rocca_2010,Boulanger_2014,Jacquemin_2015a,Bruneval_2015,Jacquemin_2015b,Hirose_2015,Jacquemin_2017a,Jacquemin_2017b,Rangel_2017,Krause_2017,Gui_2018,Blase_2018,Liu_2020,Blase_2020,Holzer_2018a,Holzer_2018b,Loos_2020e,Loos_2021,McKeon_2022}

\section{Connection between RPA and CC}
\label{sec:RPAx}
Interestingly, RPA has strong connections with coupled-cluster (CC) theory, \cite{Freeman_1977,Scuseria_2008,Jansen_2010,Scuseria_2013,Peng_2013,Berkelbach_2018,Rishi_2020} the workhorse of molecular electronic structure when one is looking for high accuracy. \cite{Cizek_1966,Paldus_1972,Crawford_2000,Piecuch_2002,Bartlett_2007,Shavitt_2009} 

In a landmark paper, Scuseria \textit{et al.} \cite{Scuseria_2008} have proven that ring CC with doubles (rCCD) is equivalent to RPA with exchange (RPAx) for the computation of the correlation energy, solidifying in the process the numerical evidences provided by Freeman many years before. \cite{Freeman_1977}
Assuming the existence of $\bX{}{-1}$ (which can be proven as long as the RPAx problem is stable \cite{Scuseria_2008}), this proof can be quickly summarized starting from the RPAx linear eigensystem
\begin{equation}
	\label{eq:RPA}
	\begin{pmatrix}
		\bA{}{}		&	\bB{}{}	\\
		-\bB{}{}	&	-\bA{}{}	\\
	\end{pmatrix}
	\cdot
	\begin{pmatrix}
		\bX{}{}	\\
		\bY{}{}	\\
	\end{pmatrix}
	=
	\begin{pmatrix}
		\bX{}{}	\\
		\bY{}{}	\\
	\end{pmatrix}
	\cdot
	\bOm{}{}
\end{equation}
from which one gets, by introducing $\bT{}{} = \bY{}{} \cdot \bX{}{-1}$,
\begin{equation}
	\begin{pmatrix}
		\bA{}{}		&	\bB{}{}	\\
		-\bB{}{}	&	-\bA{}{}	\\
	\end{pmatrix}
	\cdot
	\begin{pmatrix}
		\bI	\\
		\bT{}{}	\\
	\end{pmatrix}
	=
	\begin{pmatrix}
		\bI	\\
		\bT{}{}	\\
	\end{pmatrix}
	\cdot
	\bR{}{}
\end{equation}
where $\bR{}{} = \bX{}{} \cdot \bOm{}{} \cdot \bX{}{-1}$, or equivalently, the two following equations
\begin{subequations}
\begin{align}
	\label{eq:RPA_1}
	\bA{}{}	+ \bB{}{} \cdot \bT{}{} & = \bR{}{}
	\\
	\label{eq:RPA_2}
	-\bB{}{} - \bA{}{} \cdot \bT{}{} & = \bT{}{} \cdot \bR{}{}
\end{align}
\end{subequations}

Substituting Eq.~\eqref{eq:RPA_1} into Eq.~\eqref{eq:RPA_2} yields the following Riccati equation
\begin{equation}
	\bB{}{} + \bA{}{} \cdot \bT{}{} + \bT{}{} \cdot \bA{}{} + \bT{}{} \cdot \bB{}{} \cdot \bT{}{} = \bO
\end{equation}
that matches the rCCD amplitude (or residual) equations
\begin{multline}
\label{eq:rCCD}
	r_{ij}^{ab} 
	= \dbERI{ij}{ab} 
	+ \Delta_{ijab} t_{ij}^{ab}
	+ \sum_{kc} \dbERI{ic}{ak} t_{kj}^{cb} 
	\\
	+ \sum_{kc} \dbERI{kb}{cj} t_{ik}^{ac} 
	+ \sum_{klcd} \dbERI{kl}{cd} t_{ik}^{ac} t_{lj}^{db} = 0
\end{multline}
knowing that
\begin{subequations}
\begin{align}
	\label{eq:A_RPAx}
	A_{ia,jb} & = (\e{a}{} - \e{i}{}) \delta_{ij} \delta_{ab} + \dbERI{ib}{aj}
	\\ 
	\label{eq:B_RPAx}
	B_{ia,jb} & = \dbERI{ij}{ab}
\end{align}
\end{subequations}
where $\Delta_{ijab} = \e{a}{} + \e{b}{} - \e{i}{} - \e{j}{}$.
We assume real quantities throughout this paper, $\e{p}{}$ is the one-electron energy associated with the Hartree-Fock (HF) spinorbital $\SO{p}{}(\bx)$ and 
\begin{equation}
	\label{eq:ERI}
	\braket{pq}{rs} = \iint \SO{p}{}(\bx_1) \SO{q}{}(\bx_2) \frac{1}{\abs{\br_1 - \br_2}} \SO{r}{}(\bx_1) \SO{s}{}(\bx_2)  d\bx_1 d\bx_2
\end{equation}
are two-electron repulsion integrals, while 
\begin{equation}
	\dbERI{pq}{rs} = \braket{pq}{rs} - \braket{pq}{sr}
\end{equation}
are their anti-symmetrized versions.
The composite variable $\bx$ gathers spin and spatial ($\br$) variables.
The indices $i$, $j$, $k$, and $l$ are occupied (hole) orbitals; $a$, $b$, $c$, and $d$ are unoccupied (particle) orbitals; $p$, $q$, $r$, and $s$ indicate arbitrary orbitals; and $m$ labels single excitations or deexcitations.
In the following, $O$ and $V$ are the number of occupied and virtual spinorbitals, respectively, and $N = O + V$ is the total number.

There are various ways of computing the RPAx correlation energy, \cite{Furche_2008,Jansen_2010,Angyan_2011} but the usual plasmon (or trace) formula \cite{Sawada_1957a,Rowe_1968,Schuck_book} yields\footnote{The factor $1/4$ in Eq.~\eqref{eq:EcRPAx} is sometimes replaced by a factor $1/2$, which corresponds to a different choice for the interaction kernel. See Ref.~\onlinecite{Angyan_2011} for more details}
\begin{equation}
\label{eq:EcRPAx}
	\Ec^\text{RPAx} = \frac{1}{4} \Tr(\bOm{}{} - \bA{}{})
\end{equation}
and matches the rCCD correlation energy
\begin{equation}
	\Ec^\text{rCCD} = \frac{1}{4} \sum_{ijab} \dbERI{ij}{ab} t_{ij}^{ab} = \frac{1}{4} \Tr(\bB{}{} \cdot \bT{}{})
\end{equation}
because $\Tr(\bOm{}{} - \bA{}{}) = \Tr(\bR{}{} - \bA{}{}) = \Tr(\bB{}{} \cdot \bT{}{})$, as evidenced by Eq.~\eqref{eq:RPA_1}.
Note that, in the case of RPAx, the same expression as Eq.~\eqref{eq:EcRPAx} can be derived from the adiabatic connection fluctuation dissipation theorem \cite{Furche_2005} (ACFDT) when exchange is included in the interaction kernel. \cite{Angyan_2011}

This simple and elegant proof was subsequently extended to excitation energies by Berkelbach, \cite{Berkelbach_2018} who showed that similitudes between equation-of-motion (EOM) rCCD (EOM-rCCD) \cite{Stanton_1993} and RPAx exist when the EOM space is restricted to the 1h1p configurations and only the two-body terms are dressed by rCCD correlation (see also Ref.~\onlinecite{Rishi_2020}).

To be more specific, restricting ourselves to CCD, \ie, $\hT = \hT_2$, the elements of the 1h1p block of the EOM Hamiltonian read \cite{Stanton_1993}
\begin{equation}
	\mel*{ \Psi_{i}^{a} }{ \bHN}{ \Psi_{j}^{b} } = \cF_{ab} \delta_{ij} - \cF_{ij} \delta_{ab} + \cW_{jabi}
\end{equation}
where $\bHN = e^{-\hT} \hH_{N} e^{\hT} - \ECC $ is the (shifted) similarity-transformed normal-ordered Hamiltonian, $\Psi_{i}^{a}$ are singly-excited determinants, the one-body terms are
\begin{subequations}
\begin{align}
\label{eq:cFab}
	\cF_{ab} & = \e{a}{} \delta_{ab} - \frac{1}{2} \sum_{klc} \dbERI{kl}{bc} t_{kl}^{ac}
	\\
\label{eq:cFij}
	\cF_{ij} & = \e{i}{} \delta_{ij} + \frac{1}{2} \sum_{kcd} \dbERI{ik}{cd} t_{jk}^{cd}
\end{align}
\end{subequations}
and the two-body term is
\begin{equation}
\label{eq:cWibaj}
	\cW_{ibaj} = \dbERI{ib}{aj} + \sum_{kc} \dbERI{ik}{ac} t_{kj}^{cb}
\end{equation}
Neglecting the effect of $\hT_2$ on the one-body terms [see Eqs.~\eqref{eq:cFab} and \eqref{eq:cFij}] and relying on the rCCD amplitudes in the two-body terms, Eq.~\eqref{eq:cWibaj}, yields 
\begin{equation}
\label{eq:EOM-rCCD}
\begin{split}
	\mel*{ \Psi_{i}^{a} }{ \bHN }{ \Psi_{j}^{b} } 
	& = (\e{a}{} - \e{i}{}) \delta_{ij} \delta_{ab} + \dbERI{ib}{aj} + \sum_{kc} \dbERI{ik}{ac} t_{kj}^{cb}
	\\
	& = (\bA{}{} + \bB{}{} \cdot \bT{}{})_{ia,jb}
\end{split}
\end{equation}
which exactly matches Eq.~\eqref{eq:RPA_1}.
Although the excitation energies of this approximate EOM-rCCD scheme are equal to the RPAx ones, it has been shown that the transition amplitudes (or residues) are distinct and only agrees at the lowest order in the Coulomb interaction. \cite{Emrich_1981,Berkelbach_2018}
Equation \eqref{eq:EOM-rCCD} can be more systematically derived through the formulation of $\Lambda$ equations based on a rCCD effective Hamiltonian, as proposed by Rishi \textit{et al.} \cite{Rishi_2020}

As we shall see below, the connection between a ph eigensystem with the structure of Eq.~\eqref{eq:RPA} and a set of CC-like amplitude equations does not hold only for RPAx as it is actually quite general and can be applied to most ph problems, such as time-dependent density-functional theory (TD-DFT), \cite{Runge_1984,Casida_1995} BSE, and many others.
This analysis has also been extended to the pp and hh sectors independently by Peng \textit{et al.} \cite{Peng_2013} and Scuseria \textit{et al.} \cite{Scuseria_2013} 
(See also Ref.~\onlinecite{Berkelbach_2018} for the extension to excitation energies for the pp and hh channels.)

\section{Connection between BSE and CC}
\label{sec:BSE}
Within the usual static approximation of BSE, one must solve a very similar linear eigenvalue problem
\begin{equation}
	\label{eq:BSE}
	\begin{pmatrix}
		\bA{}{\BSE}		&	\bB{}{\BSE}	\\
		-\bB{}{\BSE}	&	-\bA{}{\BSE}	\\
	\end{pmatrix}
	\cdot
	\begin{pmatrix}
		\bX{}{\BSE}	\\
		\bY{}{\BSE}	\\
	\end{pmatrix}
	=
	\begin{pmatrix}
		\bX{}{\BSE}	\\
		\bY{}{\BSE}	\\
	\end{pmatrix}
	\cdot
	\bOm{}{\BSE}
\end{equation}
where the matrix elements read
\begin{subequations}
\begin{align}
	\label{eq:A_BSE}
	A_{ia,jb}^{\BSE} & = \delta_{ij} \delta_{ab} (\e{a}{\GW} - \e{i}{\GW}) + \dbERI{ib}{aj} - W_{ij,ba}^\text{stat}
	\\ 
	\label{eq:B_BSE}
	B_{ia,jb}^{\BSE} & = \dbERI{ij}{ab} - W_{ib,ja}^\text{stat}
\end{align}
\end{subequations}
The quasiparticle energies $\e{p}{\GW}$ are computed at the $GW$ level (see below) and 
\begin{multline}
	\label{eq:W}
	W_{pq,rs}^\text{c}(\omega) = \sum_{m} \sERI{pq}{m} \sERI{rs}{m} 
	\\
	\times \qty[ \frac{1}{\omega - \Om{m}{\dRPA} + \ii \eta} - \frac{1}{\omega + \Om{m}{\dRPA} - \ii \eta} ]
\end{multline}
are the elements of the correlation part of the dynamically-screened Coulomb potential which is set to its static limit \ie, $W_{pq,rs}^\text{stat} = W_{pq,rs}^\text{c}(\omega = 0)$. 
In Eq.~\eqref{eq:W}, $\eta$ is a positive infinitesimal, the screened two-electron integrals are
\begin{equation}
	\label{eq:sERI}
	\sERI{pq}{m} = \sum_{ia} \ERI{pi}{qa} \qty( \bX{m}{\dRPA} + \bY{m}{\dRPA} )_{ia}
\end{equation}
and $\Om{m}{\dRPA}$ is the $m$th (positive) eigenvalue and $\bX{m}{\dRPA} + \bY{m}{\dRPA}$ is constructed from the corresponding eigenvectors of the direct (\ie, without exchange) RPA (dRPA) problem defined as
\begin{equation}
	\label{eq:dRPA}
	\begin{pmatrix}
		\bA{}{\dRPA}		&	\bB{}{\dRPA}	\\
		-\bB{}{\dRPA}	&	-\bA{}{\dRPA}	\\
	\end{pmatrix}
	\cdot
	\begin{pmatrix}
		\bX{}{\dRPA}	\\
		\bY{}{\dRPA}	\\
	\end{pmatrix}
	=
	\begin{pmatrix}
		\bX{}{\dRPA}	\\
		\bY{}{\dRPA}	\\
	\end{pmatrix}
	\cdot
	\bOm{}{\dRPA}
\end{equation}
with
\begin{subequations}
\begin{align}
	\label{eq:A_dRPA}
	A_{ia,jb}^{\dRPA} & = \delta_{ij} \delta_{ab} (\e{a}{\GW} - \e{i}{\GW}) + \ERI{ib}{aj}
	\\ 
	\label{eq:B_dRPA}
	B_{ia,jb}^{\dRPA} & = \ERI{ij}{ab}
\end{align}
\end{subequations}

As readily seen in Eqs.~\eqref{eq:A_RPAx}, \eqref{eq:B_RPAx}, \eqref{eq:A_BSE} and \eqref{eq:B_BSE}, the only difference between RPAx and BSE lies in the definition of the matrix elements, where one includes, via the presence of the $GW$ quasiparticle energies in the one-body terms and the screening of the electron-electron interaction [see Eq.~\eqref{eq:W}] in the two-body terms, correlation effects at the BSE level.
Therefore, following the derivation detailed in Sec.~\ref{sec:RPAx}, one can show that the BSE correlation energy obtained using the trace formula 
\begin{equation}
	\Ec^\text{BSE} = \frac{1}{4} \Tr(\bOm{}{\BSE} - \bA{}{\BSE}) = \frac{1}{4} \sum_{ijab} \wERI{ij}{ab} \tilde{t}_{ij}^{ab} 
\end{equation}
can be equivalently obtained via a set of rCCD-like amplitude equations, where one substitutes in Eq.~\eqref{eq:rCCD} the HF orbital energies by the $GW$ quasiparticle energies and all the antisymmetrized two-electron integrals $\dbERI{pq}{rs}$ by $\wERI{pq}{rs} = \dbERI{pq}{rs} - W_{ps,qr}^\text{stat}$, \ie,
\begin{multline}
\label{eq:rCCD-BSE}
	\Tilde{r}_{ij}^{ab}
	= \wERI{ij}{ab} 
	+ \Delta_{ijab}^{\GW} \tilde{t}_{ij}^{ab}
	+ \sum_{kc} \wERI{ic}{ak} \tilde{t}_{kj}^{cb} 
	\\
	+ \sum_{kc} \wERI{kb}{cj} \tilde{t}_{ik}^{ac} 
	+ \sum_{klcd} \wERI{kl}{cd} \tilde{t}_{ik}^{ac} \tilde{t}_{lj}^{db} = 0
\end{multline}
with $\Delta_{ijab}^{\GW} = \e{a}{\GW} + \e{b}{\GW} - \e{i}{\GW} - \e{j}{\GW}$.
Similarly to the diagonalization of the eigensystem \eqref{eq:BSE}, these approximate CCD amplitude equations can be solved with $\order*{N^6}$ cost via the definition of appropriate intermediates.
As in the case of RPAx (see Sec.~\ref{sec:RPAx}), several variants of the BSE correlation energy do exist, \footnote{\alert{To the best of our knowledge, the trace (or plasmon) formula has been first introduced by Sawada \cite{Sawada_1957a} to calculate the correlation energy of the uniform electron gas as an alternative to the Gell-Mann-Brueckner formulation \cite{Gell-Mann_1957} where one integrates along the adiabatic connection path.
More precisely, the trace formula can be justified via the introduction of a quadratic Hamiltonian made of boson transition operators (quasiboson approximation). See Ref.~\onlinecite{Li_2020} for more details}.} either based on the plasmon formula \cite{Li_2020,Li_2021,DiSabatino_2021} or the ACFDT. \cite{Maggio_2016,Holzer_2018b,Loos_2020e,Berger_2021,DiSabatino_2021}

Following Berkelbach's analysis, \cite{Berkelbach_2018} one can extend the connection to excited states. 
Indeed, one can obtain an analog of the 1h1p block of the approximate EOM-rCCD Hamiltonian [see Eq.\eqref{eq:EOM-rCCD}] using the amplitudes resulting from Eq.~\eqref{eq:rCCD-BSE} as well as replacing $\bA{}{}$ and $\bB{}{}$ by their BSE counterparts, \ie,
\begin{equation}
\label{eq:EOM-rCCD-BSE}
	\mel*{ \Psi_{i}^{a} }{ \tHN }{ \Psi_{j}^{b} } 
	= (\e{a}{\GW} - \e{i}{\GW}) \delta_{ij} \delta_{ab} + \wERI{ib}{aj} + \sum_{kc} \wERI{ik}{ac} \tilde{t}_{kj}^{cb}
\end{equation}
This equation provides the same excitation energies as the conventional linear-response equations \eqref{eq:BSE}, and the corresponding $\Lambda$ equations based on the BSE effective Hamiltonian $\tHN$ can be derived following Ref.~\onlinecite{Rishi_2020}.

However, there is a significant difference with RPAx as the BSE involves $GW$ quasiparticle energies, where some of the correlation has been already dressed, while the RPAx equations only involves (undressed) one-electron orbital energies, as shown in Eq.~\eqref{eq:EOM-rCCD}.
In other words, in the spirit of the Brueckner version of CCD, \cite{Handy_1989} \alert{the $GW$ pre-treatment renormalizes the bare one-electron energies and, consequently, incorporates mosaic \cite{Scuseria_2008,Scuseria_2013} as well as additional diagrams, \cite{Lange_2018} a process named Brueckner-like dressing in Ref.~\onlinecite{Berkelbach_2018}.}

This observation evidences clear similitudes between BSE@$GW$ and the similarity-transformed EOM-CC (STEOM-CC) method introduced by Nooijen, \cite{Nooijen_1997c,Nooijen_1997b,Nooijen_1997a} where one performs a second similarity transformation to partially decouple the 1h determinants from the 2h1p ones in the ionization potential (IP) sector and the 1p determinants from the 1h2p ones in the electron affinity (EA) sector.
At the CC with singles and doubles (CCSD) level, for example, this is achieved by performing IP-EOM-CCSD \cite{Stanton_1994,Musial_2003a} (up to 2h1p) and EA-EOM-CCSD \cite{Nooijen_1995,Musial_2003b} (up to 2p1h) calculations prior to the EOM-CC treatment, which can then be reduced to the 1h1p sector thanks to this partial decoupling.
(An extended version of STEOM-CC has been proposed where the EOM treatment is pushed up to 2h2p. \cite{Nooijen_2000})
Following the same philosophy, in BSE@$GW$, one performs first a $GW$ calculation (which corresponds to an approximate and simultaneous treatment of the IP and EA sectors up to 2h1p and 2p1h \cite{Lange_2018,Monino_2022}) in order to renormalize the one-electron energies (see Sec.~\ref{sec:GW} for more details).
Then, a static BSE calculation is performed in the 1h1p sector with a two-body term dressed with correlation stemming from $GW$.
The dynamical version of BSE [where the BSE kernel is explicitly treated as frequency-dependent in Eq.~\eqref{eq:BSE}] takes partially into account the 2h2p configurations. \cite{Strinati_1980,Strinati_1982,Strinati_1984,Strinati_1988,Rohlfing_2000,Romaniello_2009b,Loos_2020h,Authier_2020,Monino_2021,Bintrim_2022} 

\section{Connection between $GW$ and CC}
\label{sec:GW}

Because $GW$ is able to capture key correlation effects as illustrated above, it is therefore interesting to investigate if it is also possible to recast the $GW$ equations as a set of CC-like equations that can be solved iteratively using the CC machinery.
Connections between approximate IP/EA-EOM-CC schemes and the $GW$ approximation have been already studied in details by Lange and Berkelbach, \cite{Lange_2018} but we believe that the present work proposes a different perspective on this particular subject as we derive genuine CC equations that do not decouple the 2h1p and 2p1h sectors.
Note also that the procedure described below can be applied to other approximate self-energies such as second-order Green's function (or second Born) \cite{Stefanucci_2013,Ortiz_2013,Phillips_2014,Rusakov_2014,Hirata_2015,Hirata_2017} or $T$-matrix.\cite{Romaniello_2012,Zhang_2017,Li_2021b,Loos_2022}

Quite unfortunately, there are several ways of computing $GW$ quasiparticle energies. \cite{Loos_2018b}
Within the perturbative $GW$ scheme (commonly known as $G_0W_0$), the quasiparticle energies are obtained via a one-shot procedure (with or without linearization).
\cite{Strinati_1980,Hybertsen_1985a,Hybertsen_1986,Godby_1988,Linden_1988,Northrup_1991,Blase_1994,Rohlfing_1995,Shishkin_2007}
Partial self-consistency can be attained via the \textit{``eigenvalue''} self-consistent $GW$ (ev$GW$)  \cite{Hybertsen_1986,Shishkin_2007,Blase_2011,Faber_2011,Rangel_2016,Gui_2018} or the quasiparticle self-consistent $GW$ (qs$GW$) \cite{Faleev_2004,vanSchilfgaarde_2006,Kotani_2007,Ke_2011,Kaplan_2016} schemes.

In the most general setting, the quasiparticle energies and their corresponding orbitals are obtained by diagonalizing the so-called non-linear and frequency-dependent quasiparticle equation 
\begin{equation}
	\label{eq:GW}
	\qty[ \be{}{} + \bSig{}{\GW}\qty(\omega = \e{p}{\GW}) ] \SO{p}{\GW} = \e{p}{\GW} \SO{p}{\GW}
\end{equation}
\alert{which gives also access to the satellite solutions}.
In Eq.~\eqref{eq:GW}, $\be{}{}$ is a diagonal matrix gathering the HF orbital energies and the elements of the correlation part of the dynamical (and non-hermitian) $GW$ self-energy are
\begin{equation}
\begin{split}
	\Sig{pq}{\GW}(\omega) 
	& = \sum_{im} \frac{\sERI{pi}{m} \sERI{qi}{m}}{\omega - \e{i}{\GW} + \Om{m}{\dRPA} - \ii \eta}
	\\
	& + \sum_{am} \frac{\sERI{pa}{m} \sERI{qa}{m}}{\omega - \e{a}{\GW} - \Om{m}{\dRPA} + \ii \eta}
\end{split}
\end{equation}
Because both the left- and right-hand sides of Eq.~\eqref{eq:GW} depend on $\e{p}{\GW}$, this equation has to be solved iteratively via a self-consistent procedure.

As shown by Bintrim and Berkelbach, \cite{Bintrim_2021} the quasiparticle equation \eqref{eq:GW} can be recast as a larger set of linear and frequency-independent equations (that still needs to be solved self-consistently), which reads in the Tamm-Dancoff approximation
\begin{equation}
\label{eq:GWlin}
	\begin{pmatrix}
		\be{}{}						&	\bV{}{\text{2h1p}}		&	\bV{}{\text{2p1h}}	\\
		\T{(\bV{}{\text{2h1p}})}	&	\bC{}{\text{2h1p}}		&	\bO					\\
		\T{(\bV{}{\text{2p1h}})}	&	\bO						&	\bC{}{\text{2p1h}}	\\
	\end{pmatrix}
	\cdot
	\begin{pmatrix}
		\bX{}{}	\\
		\bY{}{\text{2h1p}}	\\
		\bY{}{\text{2p1h}}	\\
	\end{pmatrix}
	=
	\begin{pmatrix}
		\bX{}{}	\\
		\bY{}{\text{2h1p}}	\\
		\bY{}{\text{2p1h}}	\\
	\end{pmatrix}
	\cdot
	\be{}{\GW}
\end{equation}
where $\be{}{\GW}$ is a diagonal matrix collecting the quasiparticle energies, the 2h1p and 2p1h matrix elements are
\begin{subequations}
\begin{align}
	C^\text{2h1p}_{ija,klc} & = \qty[ \qty( \e{i}{\GW} + \e{j}{\GW} - \e{a}{\GW}) \delta_{jl} \delta_{ac} - \ERI{jc}{al} ] \delta_{ik} 
	\\
	C^\text{2p1h}_{iab,kcd} & = \qty[ \qty( \e{a}{\GW} + \e{b}{\GW} - \e{i}{\GW}) \delta_{ik} \delta_{ac} + \ERI{ak}{ic} ] \delta_{bd} 
\end{align}
\end{subequations}
and the corresponding coupling blocks read
\begin{align}
	V^\text{2h1p}_{p,klc} & = \ERI{pc}{kl}
	&
	V^\text{2p1h}_{p,kcd} & = \ERI{pk}{dc}
\end{align}
Going beyond the Tamm-Dancoff approximation is possible, but more cumbersome. \cite{Bintrim_2021}
Note that, contrary to the IP/EA-EOM-CC equations, $GW$ does couple the IP and EA sectors due to the lack of exponential parametrization of the wave function. \cite{Nooijen_1995,Rishi_2020}
However, it allows to generate higher-order diagrams. \cite{Lange_2018,Schirmer_2018}

Let us suppose that we are looking for the $N$ \textit{``principal''} (\ie, quasiparticle) solutions of the eigensystem \eqref{eq:GWlin}. 
Therefore, $\bX{}{}$ and $\be{}{\GW}$ are square matrices of size $N \times N$.
Assuming the existence of $\bX{}{-1}$ and introducing $\bT{}{\text{2h1p}} = \bY{}{\text{2h1p}} \cdot \bX{}{-1}$ and $\bT{}{\text{2p1h}} = \bY{}{\text{2p1h}} \cdot \bX{}{-1}$, we have
\begin{equation}
	\begin{pmatrix}
		\be{}{}						&	\bV{}{\text{2h1p}}		&	\bV{}{\text{2p1h}}	\\
		\T{(\bV{}{\text{2h1p}})}	&	\bC{}{\text{2h1p}}		&	\bO					\\
		\T{(\bV{}{\text{2p1h}})}	&	\bO						&	\bC{}{\text{2p1h}}	\\
	\end{pmatrix}
	\cdot
	\begin{pmatrix}
		\bI	\\
		\bT{}{\text{2h1p}}	\\
		\bT{}{\text{2p1h}}	\\
	\end{pmatrix}
	=
	\begin{pmatrix}
		\bI	\\
		\bT{}{\text{2h1p}}	\\
		\bT{}{\text{2p1h}}	\\
	\end{pmatrix}
	\cdot
	\bR{}{}
\end{equation}
with $\bR{}{} = \bX{}{} \cdot \be{}{\GW} \cdot \bX{}{-1}$, which yields the three following equations
\begin{subequations}
\begin{align}
	\be{}{} + \bV{}{\text{2h1p}} \cdot \bT{}{\text{2h1p}} + \bV{}{\text{2p1h}} \cdot \bT{}{\text{2p1h}} & = \bR{}{}
	\label{eq:R}
	\\
	\T{(\bV{}{\text{2h1p}})} + \bC{}{\text{2h1p}} \cdot \bT{}{\text{2h1p}} & = \bT{}{\text{2h1p}} \cdot \bR{}{}
	\label{eq:T1R}
	\\
	\T{(\bV{}{\text{2p1h}})} + \bC{}{\text{2p1h}} \cdot \bT{}{\text{2p1h}} & = \bT{}{\text{2p1h}} \cdot \bR{}{}
	\label{eq:T2R}
\end{align}
\end{subequations}
Substituting Eq.~\eqref{eq:R} into Eqs.~\eqref{eq:T1R} and \eqref{eq:T2R}, one gets two coupled Riccati equations 
\begin{subequations}
\begin{align}
	\begin{split}
		\T{(\bV{}{\text{2h1p}})} 
		+ \bC{}{\text{2h1p}} \cdot \bT{}{\text{2h1p}} 
		- \bT{}{\text{2h1p}} \cdot \be{}{} 
		- \bT{}{\text{2h1p}} \cdot \bV{}{\text{2h1p}} \cdot \bT{}{\text{2h1p}}
		\\
		- \bT{}{\text{2h1p}} \cdot \bV{}{\text{2p1h}} \cdot \bT{}{\text{2p1h}}
		= \bO
	\end{split}
	\\
	\begin{split}
		\T{(\bV{}{\text{2p1h}})} 
		+ \bC{}{\text{2p1h}} \cdot \bT{}{\text{2p1h}} 
		- \bT{}{\text{2p1h}} \cdot \be{}{} 
		- \bT{}{\text{2p1h}} \cdot \bV{}{\text{2h1p}} \cdot \bT{}{\text{2h1p}}
		\\
		- \bT{}{\text{2p1h}} \cdot \bV{}{\text{2p1h}} \cdot \bT{}{\text{2p1h}}
		= \bO
	\end{split}
\end{align}
\end{subequations}
that can be converted to the following CC-like residual equations
\begin{subequations}
\begin{align}
	\label{eq:r_2h1p}
	\begin{split}
		r_{ija,p}^{\text{2h1p}}
		& = \ERI{pa}{ij}
		+ \Delta_{ija,p}^{\text{2h1p}} t_{ija,p}^{\text{2h1p}} 
		- \sum_{kc} \ERI{jc}{ak} t_{ikc,p}^{\text{2h1p}} 
		\\
		& - \sum_{klcq} \ERI{qc}{kl} t_{ija,q}^{\text{2h1p}} t_{klc,p}^{\text{2h1p}} 
		- \sum_{kcdq} \ERI{qk}{dc} t_{ija,q}^{\text{2h1p}} t_{kcd,p}^{\text{2p1h}} 
		= 0
	\end{split}
	\\
	\label{eq:r_2p1h}
	\begin{split}
		r_{iab,p}^{\text{2p1h}}
		& = \ERI{pi}{ba}
		+ \Delta_{iab,p}^{\text{2p1h}} t_{iab,p}^{\text{2p1h}} 
		+ \sum_{kc} \ERI{ak}{ic} t_{kcb,p}^{\text{2p1h}} 
		\\
		& - \sum_{klcq} \ERI{qc}{kl} t_{iab,q}^{\text{2p1h}} t_{klc,p}^{\text{2h1p}} 
		- \sum_{kcdq} \ERI{qk}{dc} t_{iab,q}^{\text{2p1h}} t_{kcd,p}^{\text{2p1h}} 
		= 0
	\end{split}
\end{align}
\end{subequations}
with $\Delta_{ija,p}^{\text{2h1p}} = \e{i}{\GW} + \e{j}{\GW} - \e{a}{\GW} - \e{p}{}$ and $\Delta_{iab,p}^{\text{2p1h}} = \e{a}{\GW} + \e{b}{\GW} - \e{i}{\GW} - \e{p}{}$.
To determine the 2h1p and 2p1h amplitudes, $t_{ija,p}^{\text{2h1p}}$ and $t_{iab,p}^{\text{2p1h}} $, one can then rely on the usual quasi-Newton iterative procedure to solve these quadratic equations by updating the amplitudes via
\begin{subequations}
\begin{align}
	\label{eq:t_2h1p_update}
	t_{ija,p}^{\text{2h1p}} & \leftarrow t_{ija,p}^{\text{2h1p}} - \qty( \Delta_{ija,p}^{\text{2h1p}} )^{-1} r_{ija,p}^{\text{2h1p}}
	\\
	\label{eq:t_2p1h_update}
	t_{iab,p}^{\text{2p1h}} & \leftarrow t_{iab,p}^{\text{2p1h}} - \qty( \Delta_{iab,p}^{\text{2p1h}} )^{-1} r_{iab,p}^{\text{2p1h}}
\end{align}
\end{subequations}

The quasiparticle energies $\e{p}{GW}$ are thus provided by the eigenvalues of $\be{}{} + \bSig{}{\GW}$, where
\begin{equation}
	\bSig{}{\GW} = \bV{}{\text{2h1p}} \cdot \bT{}{\text{2h1p}} + \bV{}{\text{2p1h}} \cdot \bT{}{\text{2p1h}} 
\end{equation}
\alert{Due to the non-linear nature of these equations, the iterative procedure proposed in Eqs.~\eqref{eq:t_2h1p_update} and \eqref{eq:t_2p1h_update} can potentially converge to satellite solutions. 
This is also the case at the CC level when one relies on more elaborated algorithms to converge the amplitude equations to higher-energy solutions. \cite{Piecuch_2000,Mayhall_2010,Lee_2019,Kossoski_2021,Marie_2021b}}

Again, similarly to the dynamical equations defined in Eq.~\eqref{eq:GW} which requires the diagonalization of the dRPA eigenproblem [see Eq.~\eqref{eq:dRPA}], the CC equations reported in Eqs.~\eqref{eq:r_2h1p} and \eqref{eq:r_2p1h} can be solved with $\order*{N^6}$ cost by defining judicious intermediates.
Cholesky decomposition, density fitting, and other related techniques may be employed to further reduce this scaling as it is done in conventional $GW$ calculations. \cite{Bintrim_2021,Forster_2020,Forster_2021,Duchemin_2019,Duchemin_2020,Duchemin_2021}
The $G_0W_0$ quasiparticle energies can be easily obtained via the procedure described in Ref.~\onlinecite{Monino_2022} by solving the previous equations for each value of $p$ separately.

\section{Conclusion}
Here, we have unveiled exact similarities between CC and many-body perturbation theory at the ground- and excited-state levels.
More specifically, we have shown how to recast $GW$ and BSE as non-linear CC-like equations that can be solved with the usual CC machinery at the same computational cost.
The conventional and CC-based versions of the BSE and $GW$ schemes that we have described in the present work have been implemented in the electronic structure package QuAcK \cite{QuAcK} (available at \url{https://github.com/pfloos/QuAcK}) with which we have numerically checked these exact equivalences.
Similitudes between BSE@$GW$ and STEOM-CC have been also highlighted, and may explain the reliability of BSE@$GW$ for the computation of optical excitations in molecular systems. 

We hope that the present work may provide a consistent approach for the computation of ground- and excited-state properties (such as nuclear gradients) within the $GW$ \cite{Lazzeri_2008,Faber_2011b,Yin_2013,Montserrat_2016,Zhenglu_2019} and BSE \cite{IsmailBeigi_2003,Caylak_2021,Knysh_2022} frameworks, hence broadening the applicability of these formalisms in computational photochemistry.
However, several challenges lie ahead as one must derive, for example, the $\Lambda$ equations associated with $GW$\cite{Bartlett_1986,Rishi_2020} and the response of the static screening with respect to the external perturbation at the BSE level.
The present connections between CC and $GW$ could also provide new directions for the development of multireference $GW$ methods \cite{Brouder_2009,Linner_2019} in order to treat strongly correlated systems. \cite{Lyakh_2012,Evangelista_2018}


\acknowledgments{
PFL thanks Xavier Blase, Pina Romaniello, and Francesco Evangelista for useful discussions.
This project has received funding from the European Research Council (ERC) under the European Union's Horizon 2020 research and innovation programme (Grant agreement No.~863481).}

\section*{Data availability statement}
Data sharing is not applicable to this article as no new data were created or analyzed in this study.

%

\end{document}